\documentclass[12pt,twoside,a4paper]{article}
\usepackage{etex}
\usepackage[T1]{fontenc}
\usepackage[latin1]{inputenc}
\usepackage[
  up,
  centerlast,
  hang,
  font=small,
  format=plain,
  labelfont=bf,
  textfont=normal,
  justification=justified,
  singlelinecheck=false]{caption}
\usepackage[sumlimits,intlimits,namelimits]{amsmath}
\usepackage{amssymb,amsfonts}         
\usepackage{amsbsy,amsxtra}
\usepackage{bbold,isomath,mathdots}
\usepackage{bbm}
\usepackage{bm}
\usepackage{stmaryrd,wasysym}
\usepackage{dsfont,latexsym}
\usepackage{mathtools}
\usepackage{mathrsfs}
\usepackage{commath}                  
\usepackage{empheq}
\usepackage{braket}                   
\usepackage{eurosym}
\usepackage{color}                    
\usepackage{hyperref}                 
\usepackage{graphicx}
\usepackage{grffile}
\usepackage{afterpage,lastpage,fancyhdr}
\usepackage{authblk}
\usepackage{microtype}
\usepackage{tabularx}
\usepackage{subcaption} 
\usepackage{booktabs}
\usepackage{feynmp}
\usepackage{multicol}
\usepackage{color} 
\usepackage{array}
\usepackage{xspace}
\xspaceaddexceptions{]}
\usepackage[stable]{footmisc}  
\newcommand{\tprim}{\ensuremath{t'}\xspace}

\newcommand{\refl}{\ensuremath{\varepsilon}\xspace}

\newcommand{\GEV}{\ensuremath{\,\text{GeV}/c^2}\xspace}
\newcommand{\GEVCA}{\ensuremath{\,\text{GeV}/c}\xspace}
\newcommand{\GEVC}{\ensuremath{\,(\text{GeV}/c)^2}\xspace}

\newcommand{\MEV}{\ensuremath{\,\text{MeV}/c^2}\xspace}
\newcommand{\MPPP}{\ensuremath{m_{3\pi}}\xspace}

\newcommand{\PIPIS}{\ensuremath{[\pi\pi]_S}\xspace}

\newcommand{\PIPIPI}{\ensuremath{\pi^-\pi^-\pi^+}\xspace}

\newcommand{\Reaction}{\ensuremath{\pi^- + p \to \PIPIPI + p_\text{recoil}}\xspace}

\begin{document}
\title{Precision spectroscopy with COMPASS and the observation of a new iso-vector resonance}
%
%

\author{Stephan Paul \\
{\small 
	on behalf of
        the COMPASS collaboration
} 
    \\Physik Department E18, Technische Universit\"at M\"unchen, \\James Franck Str., D-85748 \\and
\\Excellencecluster "Universe", Boltzmannstr. 2, D-85748 Garching
}

\maketitle

\begin{abstract}
 We report on the results of a novel partial-wave analysis based on $50\cdot 10^6$ events from the reaction \Reaction at 190 $\GEVCA$ incoming beam momentum using the COMPASS spectrometer. A separated analysis in bins of $m_{3\pi}$ and four-momentum transfer $t'$ reveals the interference of resonant and non-resonant particle production and allows their spectral separation. Besides well known resonances we observe a new iso-vector meson $a_1(1420)$ at a mass of 1420 \MEV in the $f_0(980)\pi$ final state only, the origin of which is unclear. We have also examined the structure of the $0^{++}$ $\pi\pi$-isobar in the $J^{PC}=0^{-+}, 1^{++}, 2^{-+}$ three pion waves. This clearly reveals the various $0^{++}$ $\pi\pi$-isobar components and its correlation to the decay of light mesons.
\end{abstract}
%
%
\section{Introduction}
\label{intro}
Light meson spectroscopy has been performed for about 50 years using various tools and production mechanisms, each one possessing its own virtue and sensitivity to particular quantum numbers. Breakthroughs have either been obtained when new mechanisms were opened (e.g. $\overline{p}p$ annihilation \cite{CB96} or $J/\psi$ decays \cite{BES}) or when the wealth of data allowed new analysis tools to be developed (see e.g. \cite{CM0}). The COMPASS experiment is a modern high-rate spectrometer which allows three different mechanisms to be probed, diffractive-, central-  as well as quasi-real photo-production using the Primakoff reaction. Various beams are being used but in this work we focus on diffractive pion dissociation leading to a $3\pi$ final state. This reaction populates iso-vector states with all possible quantum numbers $J^{P}$, but limited to $C=+1$. With 10 to100 times more events as compared to previous works we gain new sensitivity to states with very small production cross section, open up the possibility to study our reactions in terms of $t'$ as kinematic variable and take a look inside one of the prominent isobars, namely the $\pi\pi$ S-wave. 
\section{Analysis Scheme}
\label{sec-1}
\begin{figure*}[!htb]
  \begin{center}
      \includegraphics[width=0.21\textwidth]{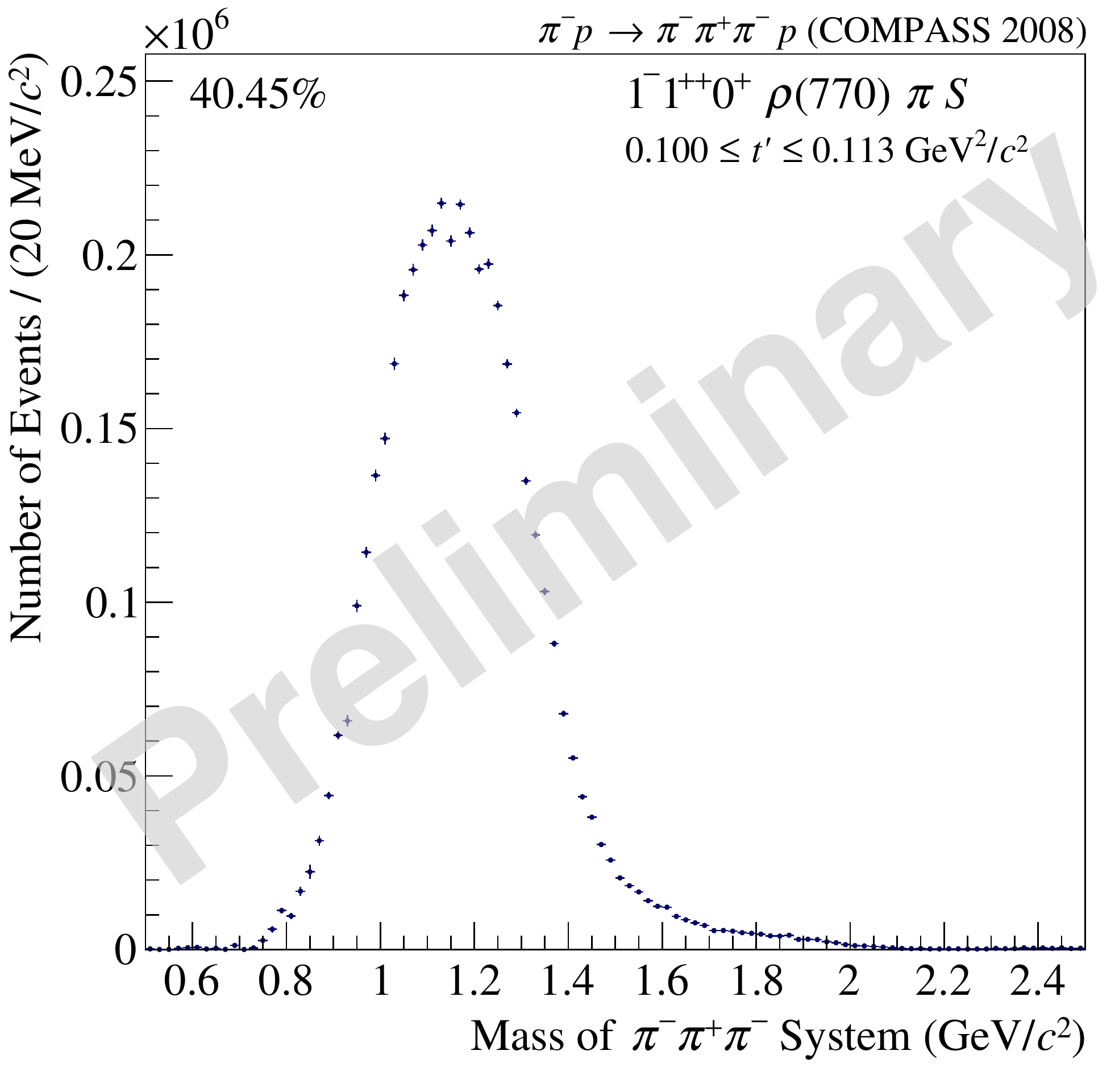}
      \includegraphics[width=0.21\textwidth]{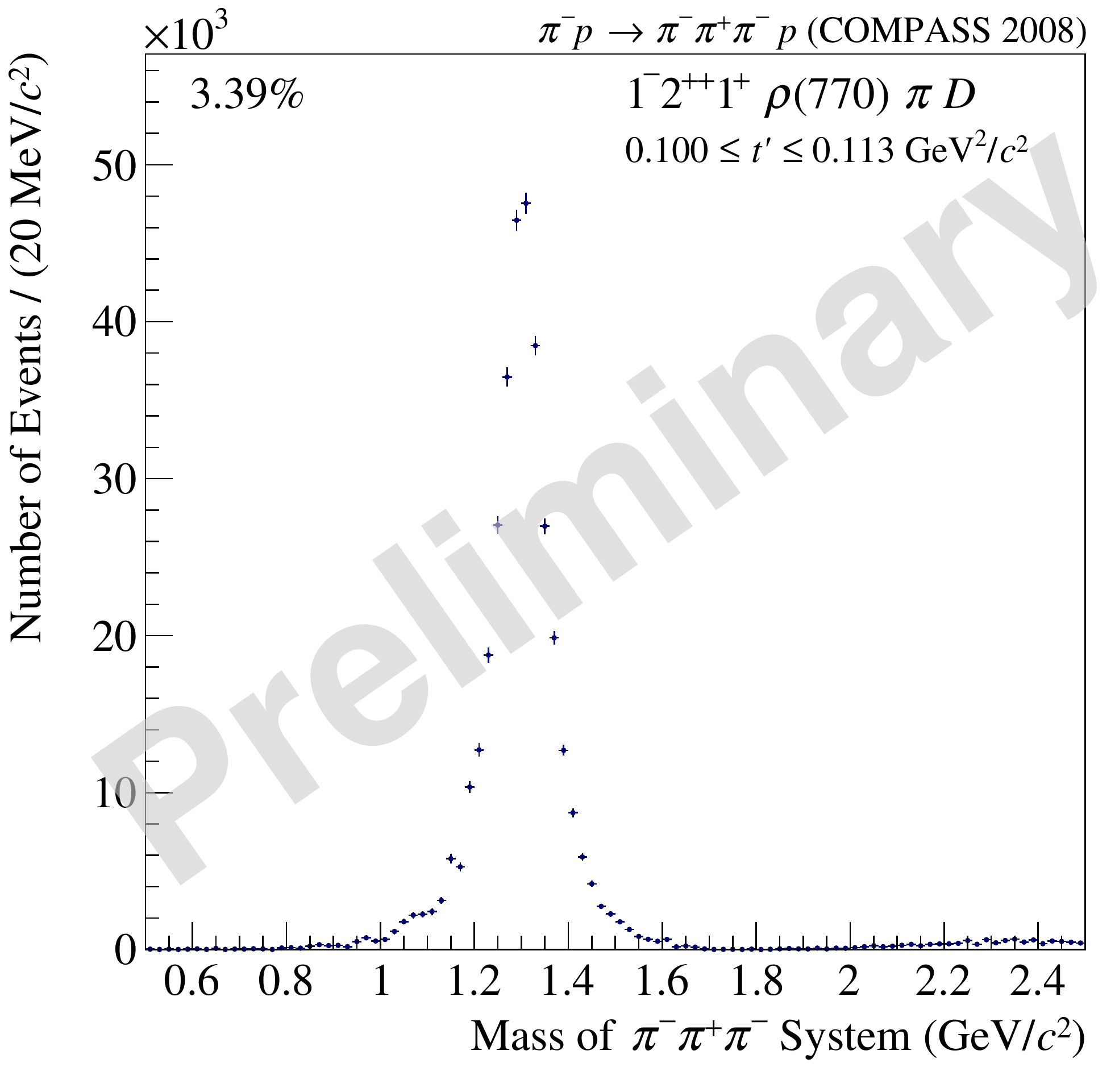}
      \includegraphics[width=0.21\textwidth]{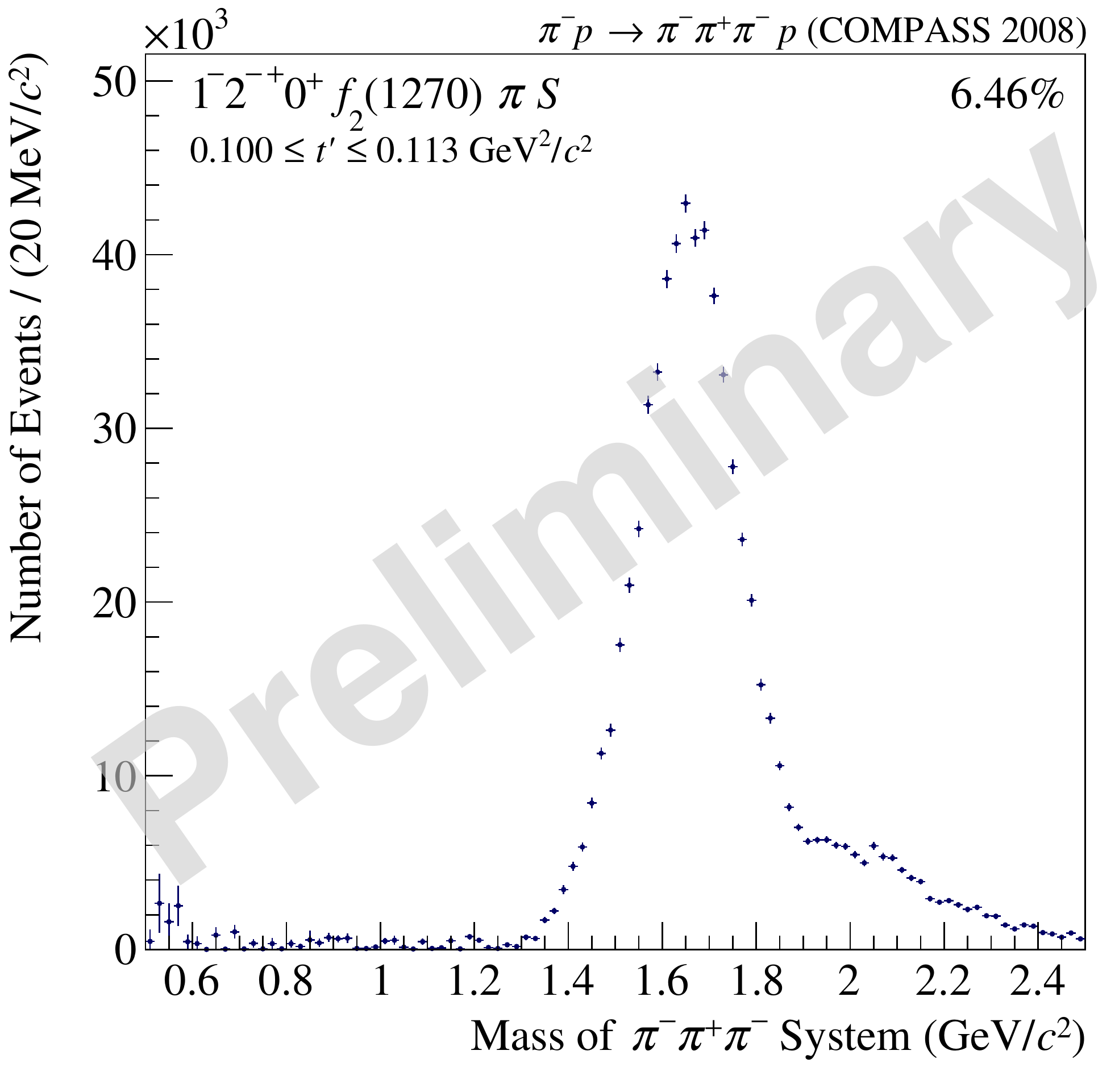}
      \includegraphics[width=0.21\textwidth]{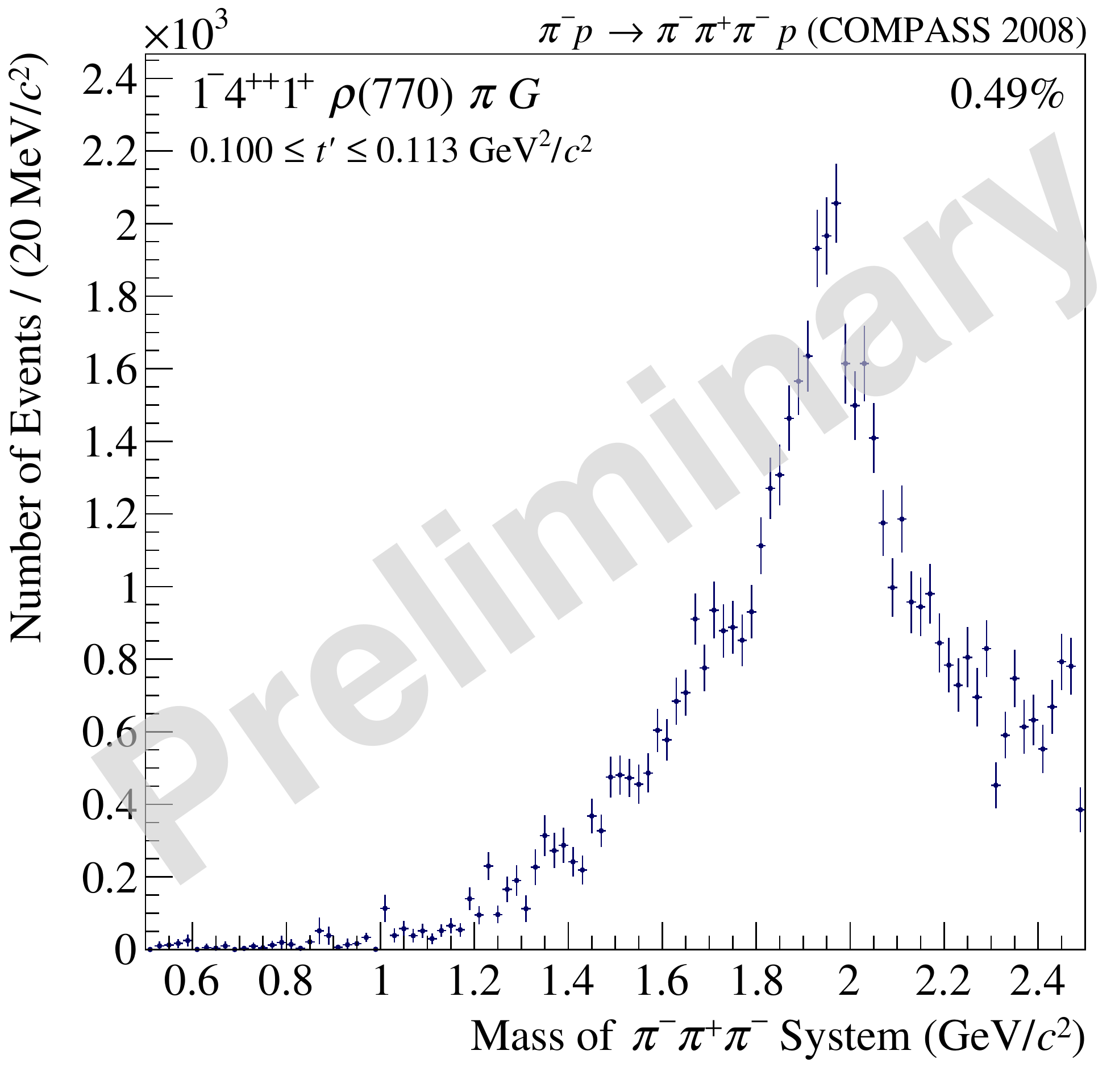}
    \\
      \includegraphics[width=0.21\textwidth]{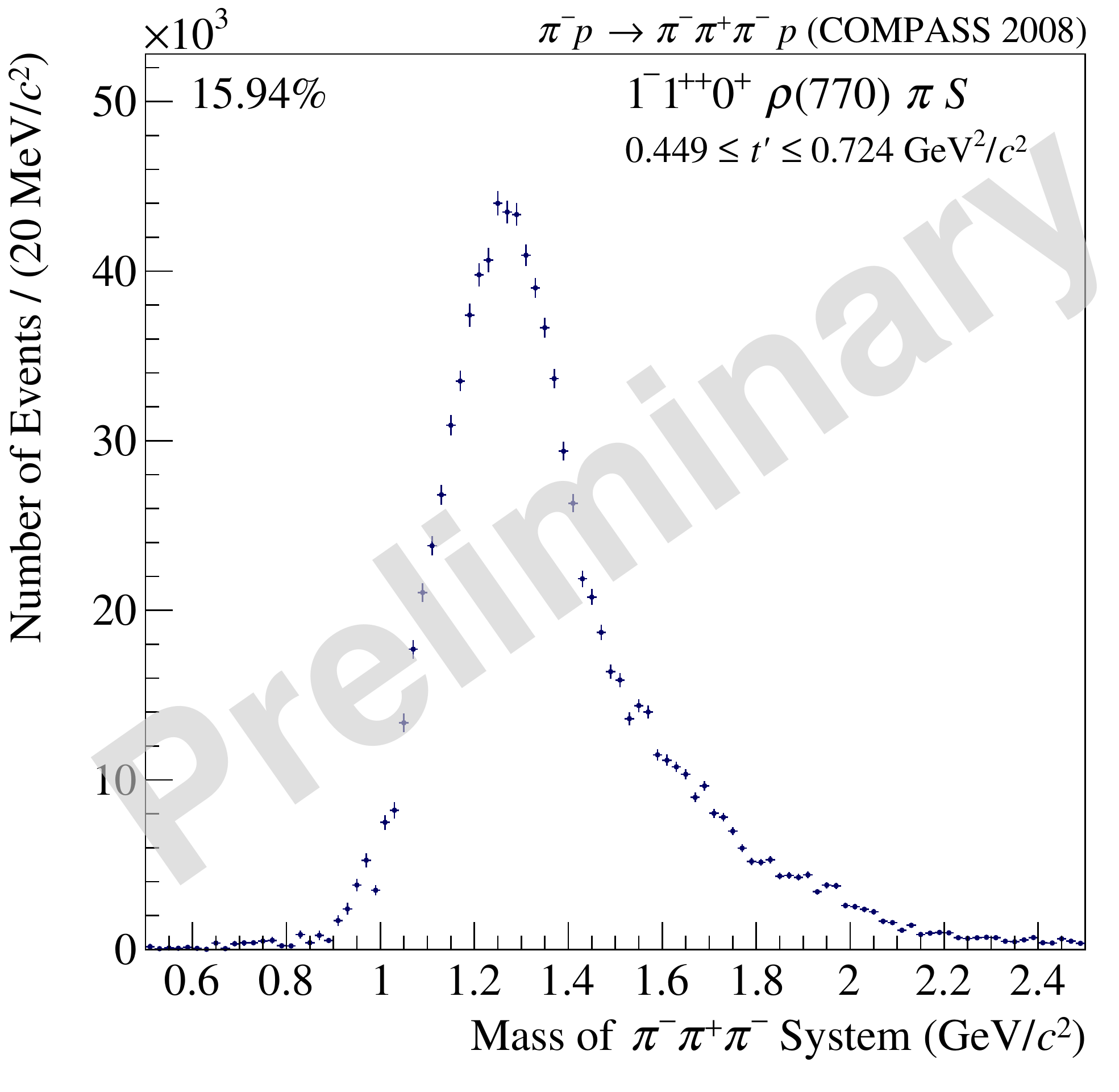}
      \includegraphics[width=0.21\textwidth]{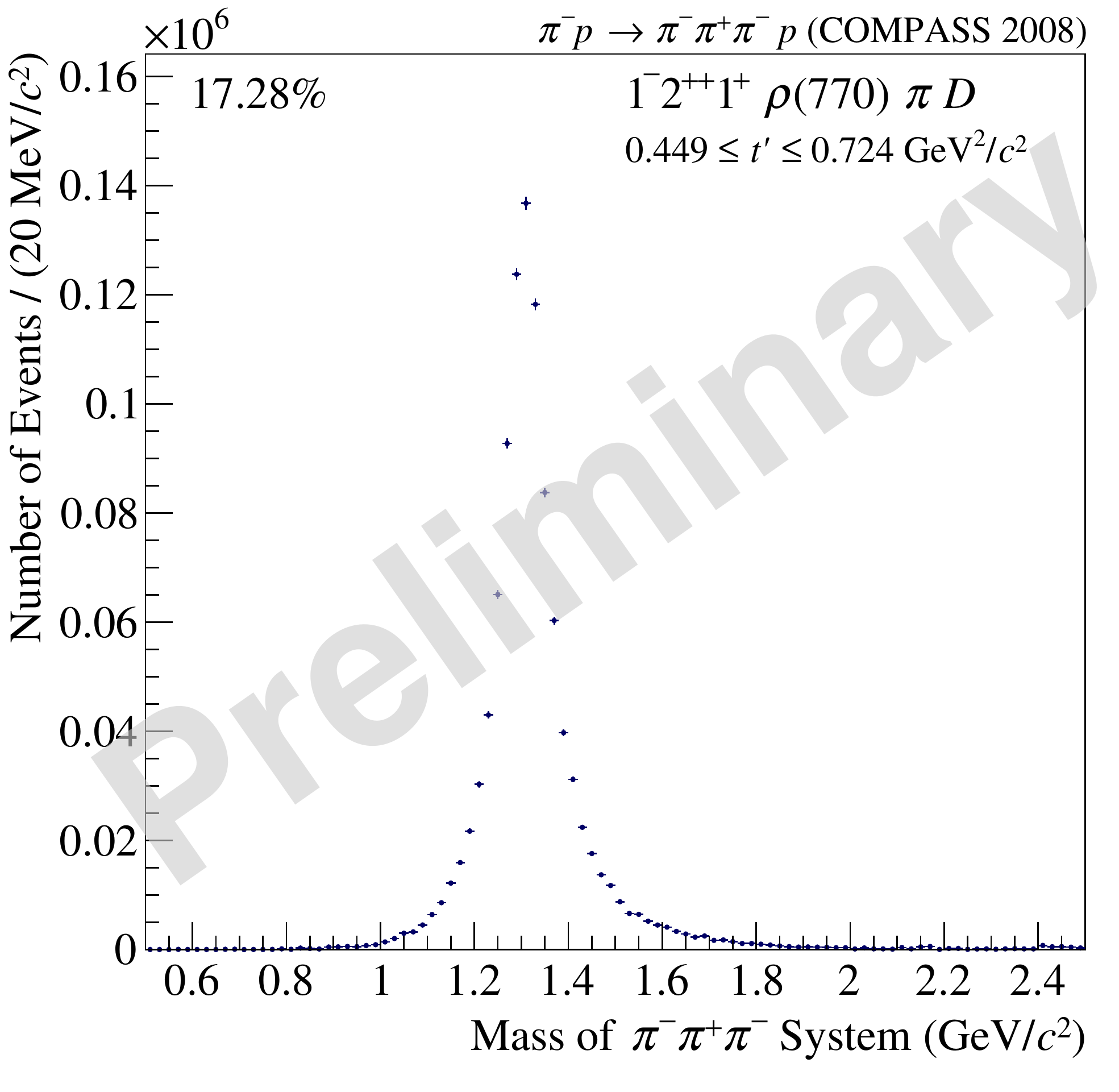}
      \includegraphics[width=0.21\textwidth]{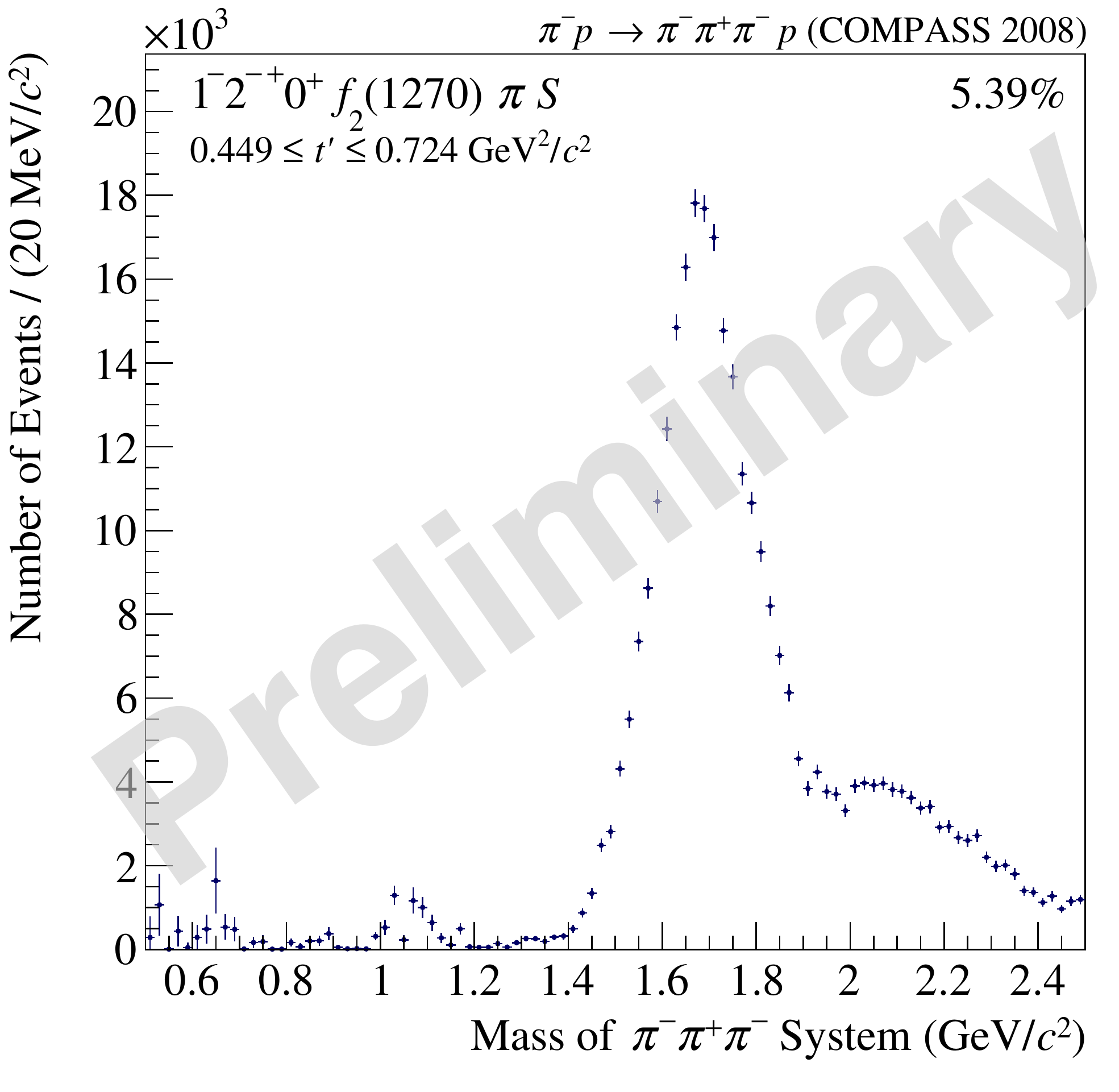}
      \includegraphics[width=0.21\textwidth]{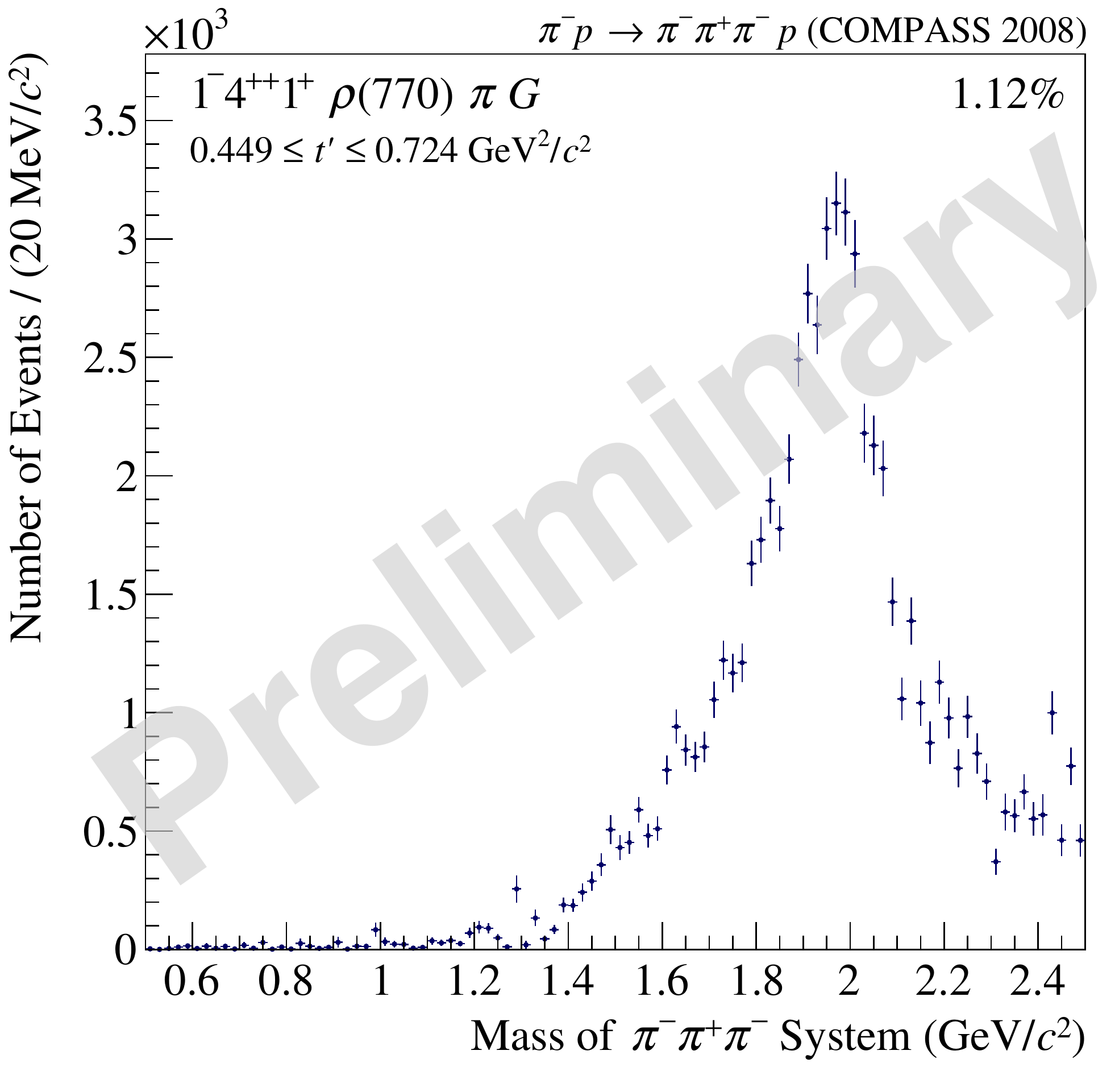}
    \caption{Intensities of four major waves in two
      different \tprim regions. Upper row: $0.100 \leq \tprim \leq
      0.113$\GEVC; lower row: $0.449 \leq \tprim \leq 0.724$\GEVC.}
    \label{fig:major_waves_t_bins}
  \end{center}
\end{figure*}

The COMPASS experiment \cite{COMPASS_Expt} is a two-stage magnetic spectrometer, covering a large solid angle, with precision vertexing and partial particle identification for both, incoming and outgoing particles. An incoming $\pi^-$-beam impinges on a $40~\text{cm}$-long $LH_2$-target surrounded by a two-layer proton recoil detector. The trigger requires a signal from this detector and vetoes events with a forward going particle being within the beam region, thus imposing a minimum value for the 4-momentum transfer $t'=|t|-|t_{min}|\approx t$ of 0.08 \GEVC. The selection of exclusive $3\pi$ final states requires a matching of incoming and outgoing momenta as well as transverse momentum balance using the recoil proton. We are left with about $50\cdot 10^6$ events which cover a mass range up to about 3 \GEV and extend to large values of $t'$ above $1 \GEVC$, although the present analysis limits itself to $0.1\leq t' \leq 1.0$ \GEVC. The analysis follows the usual scheme of a two-step partial-wave analysis \cite{Compass_pi1} where at first all events are subject to a series of fits in bins of $m_{3\pi}$ of 20 \MEV width and in 11 bins of $t'$ ({\it mass-independent fit}), chosen to equal statistics for each $t'$-bin. The PWA model is based on a sequential two-step decay into a $\pi^+\pi^-$-isobar and a bachelor $\pi^-$ with subsequent decay of the isobar. The population of the 5-dimensional phase space is modeled with a set of 88 {\it waves} each being characterized by an assumed $J^{PC}$ of the $3\pi$-state, the properties of the isobar, angular momentum $L$ enclosed by $\pi^-$ and isobar as well as magnetic quantum number $M$ and reflectivity $\varepsilon$, which is connected to the naturality of the exchange. The model allows for seven waves with $\varepsilon=-1$  and one {\it flat-wave} representing pure phase space. Both sets add incoherently to the coherent sum of 80 waves with $\varepsilon=+1$. Only waves with negligible population have been omitted from the fits presented here.
\begin{figure*}[htb]
  \begin{center}
    \includegraphics[width=0.28\textwidth]{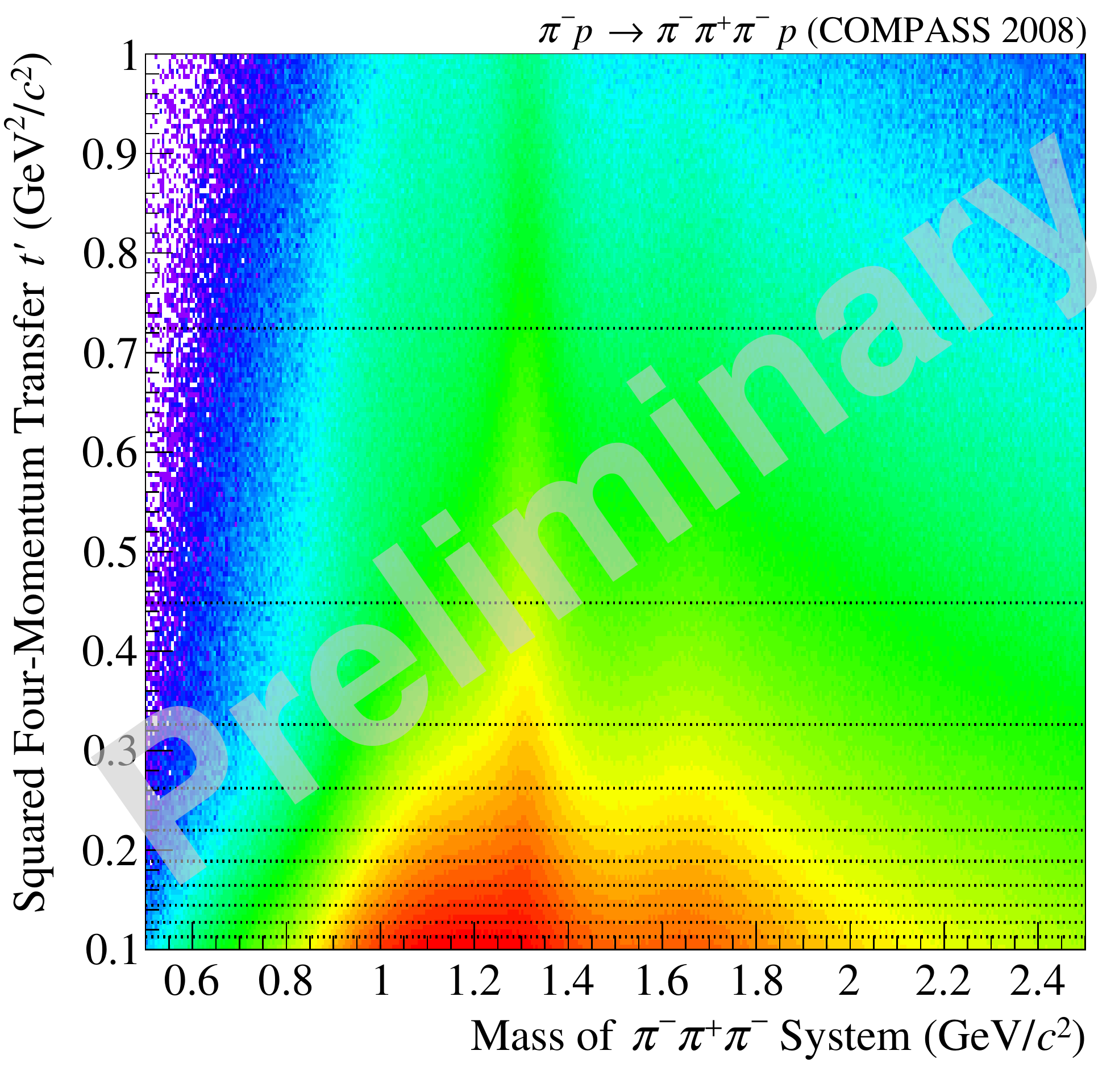}
      \includegraphics[width=0.28\textwidth]{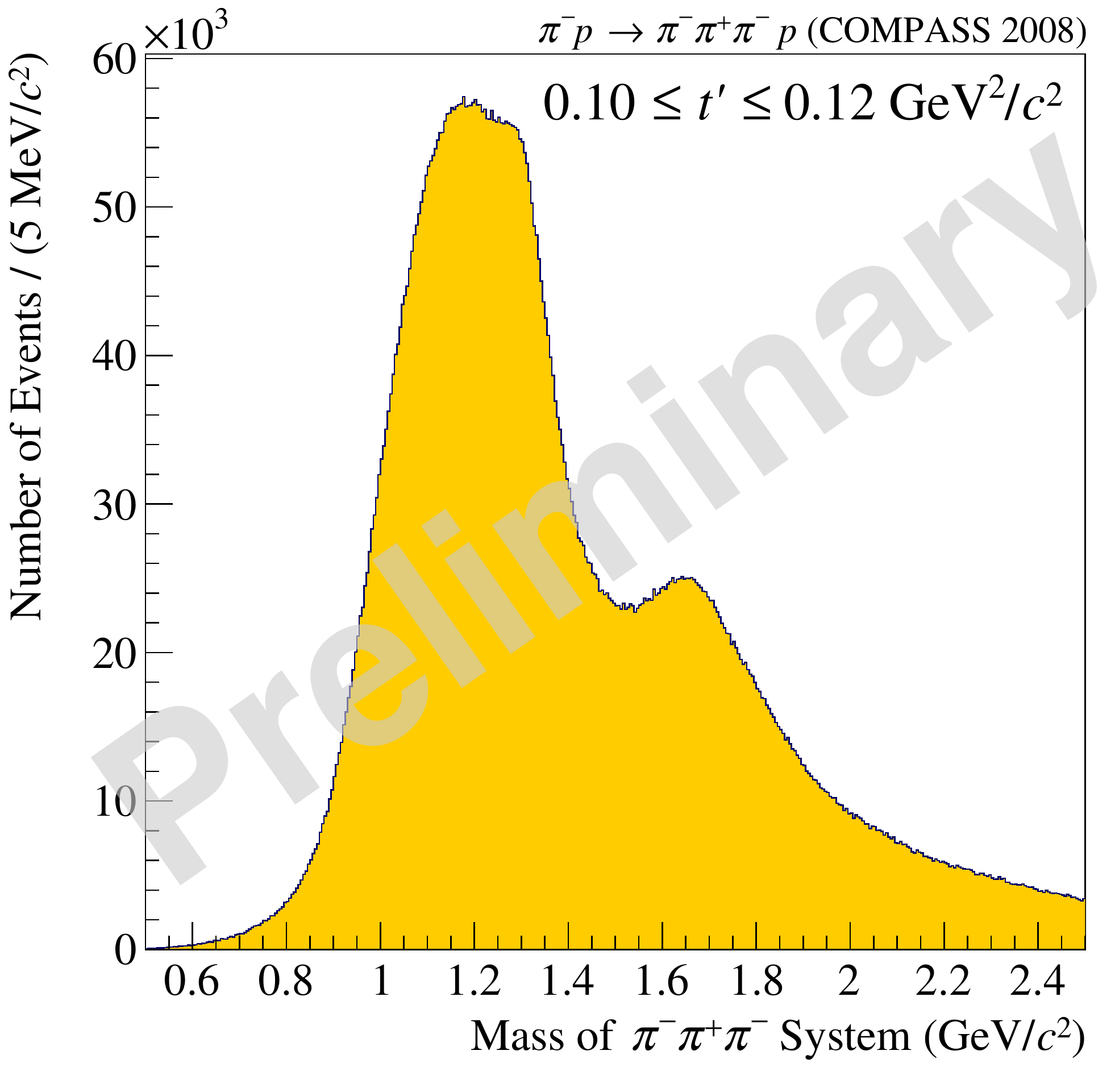}
      \includegraphics[width=0.28\textwidth]{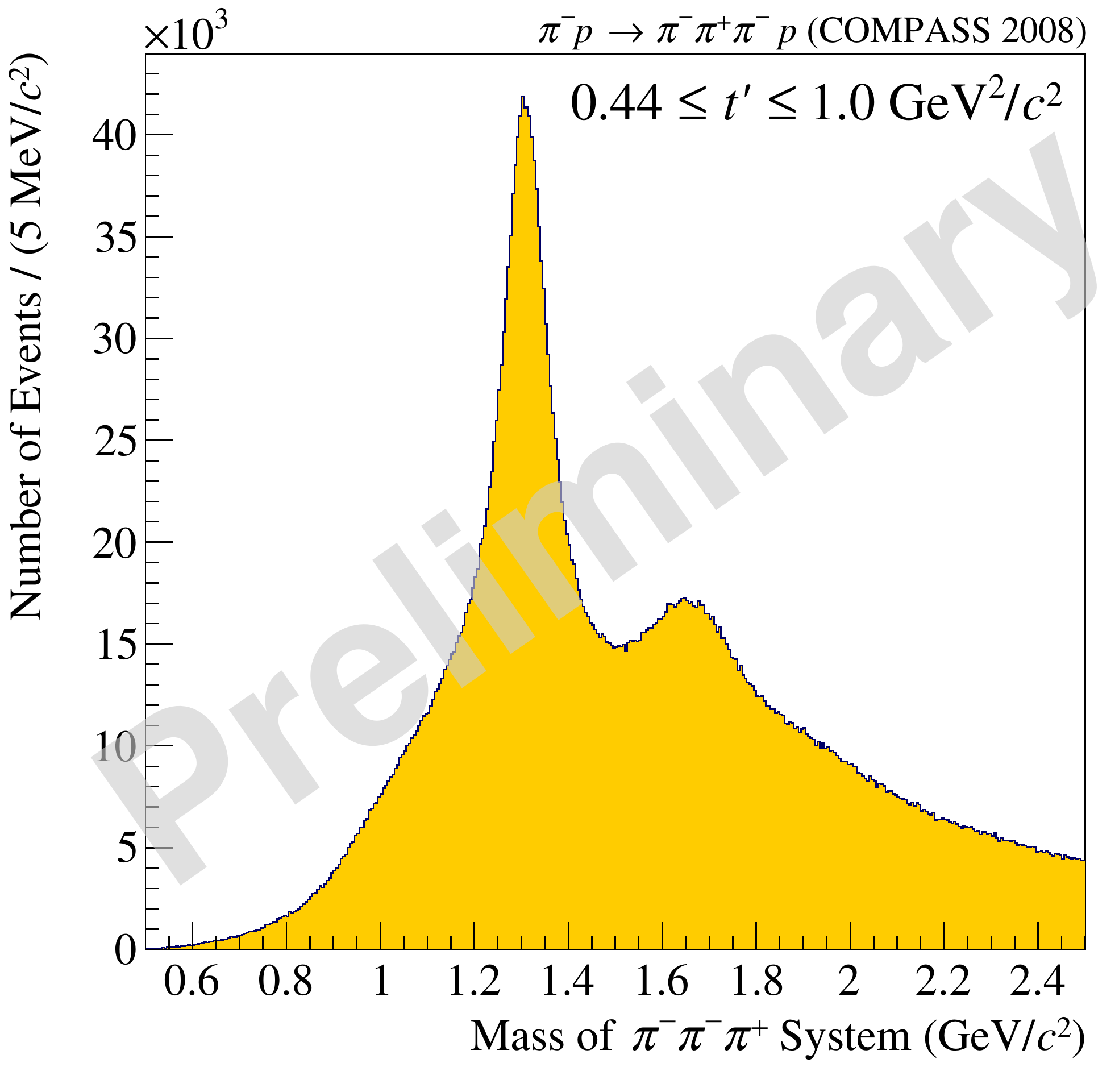}
   \caption{Left: Correlation of the squared four-momentum
      transfer \tprim and the invariant mass of the $3\pi$-system ($z$
      axis in log scale). The partial-wave analysis is performed in
      20\MEV wide bins of $m_{3\pi}$ as well as in bins of \tprim
      indicated by the horizontal lines. Centre: $m_{3\pi}$ for low values of
       \tprim ($0.10 \leq \tprim \leq 0.12$\GEVC). Right: $m_{3\pi}$ for high values of \tprim ($0.44 \leq \tprim \leq 1.00$\GEVC).}
    \label{fig:t_vs_m_binning}
\end{center}
\end{figure*}
The basic assumptions of this partial-wave analysis are founded on the observations shown in fig.~\ref{fig:t_vs_m_binning}, depicting the $3\pi$-mass spectra at low and high values of $t'$ and the correlation of $m_{3\pi}$ and $t'$, clearly demonstrating the need for a separation of these two variables. The power of this scheme, which can now be exploited in all its beauty for the first time (but had already been addressed by \cite{CM0} and \cite{BNL2}) can also be derived from fig.~\ref{fig:major_waves_t_bins}, where we depict the spectral intensity of four different but characteristic partial waves, namely $J^{PC}M^{\refl}({\it isobar})\pi L=1^{++}0^+\rho\pi S$, $2^{++}1^+\rho\pi D$, $4^{++}1^+\rho\pi G$ and $2^{-+}0^+f_2(1270)\pi S$ for two very different intervals of $t'$, dubbed {\it low} and {\it high} $t'$. 
While we clearly observe the well established states $a_2(1320)$ and $a_4(2040)$ with little change of spectral shape at different $t'$, the structures observed around the $a_1(1260)$ and $\pi_2(1670)$ reveal underlying dynamics resulting in a $t'$-dependent shape which cannot be attributed solely to a resonance. As we will see, these issues will be resolved in the second step of our PWA using a mass-dependent fit.

\section{Mass-Dependent Fits}
\label{sec-2}
\begin{figure}[!h]
	\centering
	\includegraphics[width=0.75\textwidth]{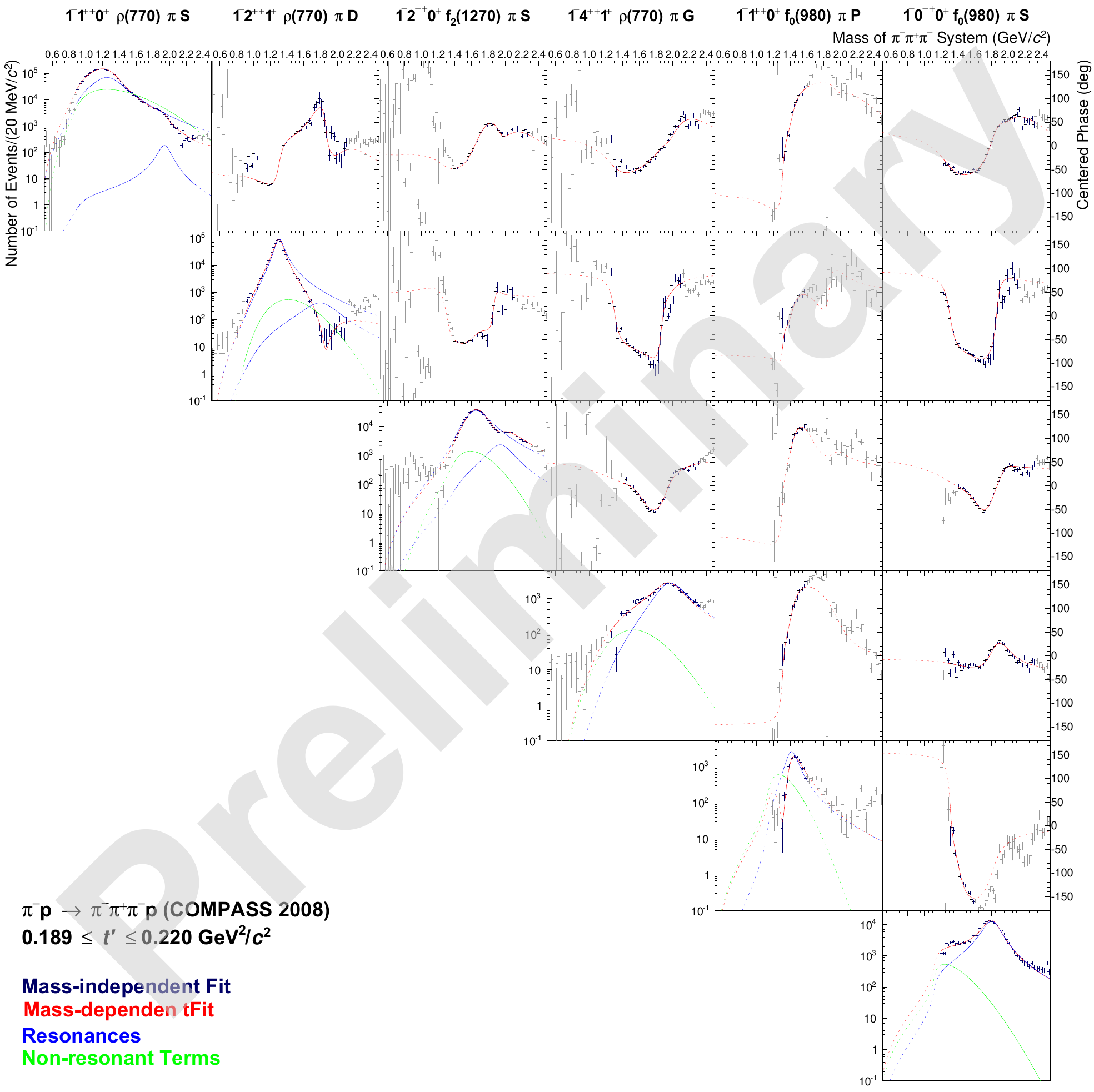}
	\caption{Result of the mass-dependent fit of the spin-density matrix for six waves
    for $t' \in [0.189, 0.220]  \GEVC$. Rows and columns correspond
    to the waves depicted in the figure. Distributions on the diagonal
    show intensities of individual waves, relative phases for pairs of
    waves are depicted on the off-diagonal. The scheme for the phase
    differences is $\phi_\text{column} - \phi_\text{row}$. For better
    visibility we have shifted all phases such that they lie in the range
    $[-180^\circ,+180^\circ]$. The red line represents the result of
    the mass-dependent fit which is performed using all data points in
    dark blue. Unused data points are drawn in grey. Green:
    non-resonant contributions; Blue: resonant contributions. Both, $a_1$ and $a_2$ decaying into $\rho\pi$ exhibit signatures for excited states around 1.8-1.9 \GEV, once appearing as a small peak, once as dip due destructive interference with a non-resonant contribution.}
	\label{fig:stamps_mid-t}
\end{figure}

\begin{figure}[!h]

\vspace{0.2cm}

  \centering
    \includegraphics[width=0.8\textwidth]{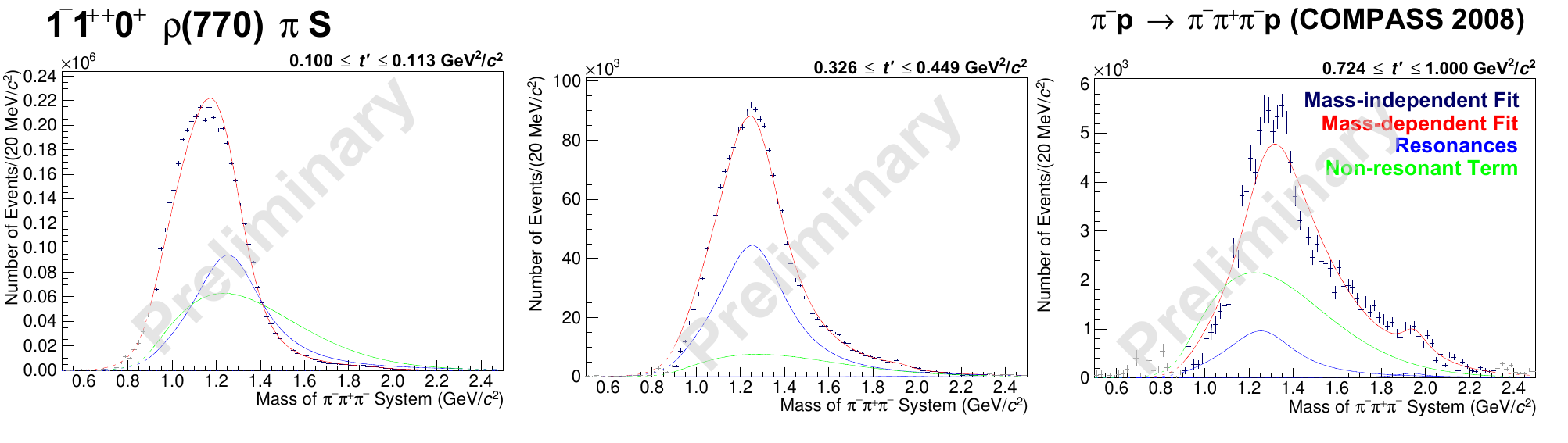}
      
  \vspace{-0.3cm}
    
  \caption{$1^{++}0^+ \rho \pi S$ intensities in three selected slices of $t'$ together with the model curve (red). Model components are shown as colored curves: non-resonant contribution (green) and $a_1(1260)$ component (blue).}
  \label{fig:1pp_rho_m0_3t}
  
\vspace{0.5cm}

\end{figure}

\begin{figure*}[!h]
  \begin{center}
        \includegraphics[width=0.345\textwidth]{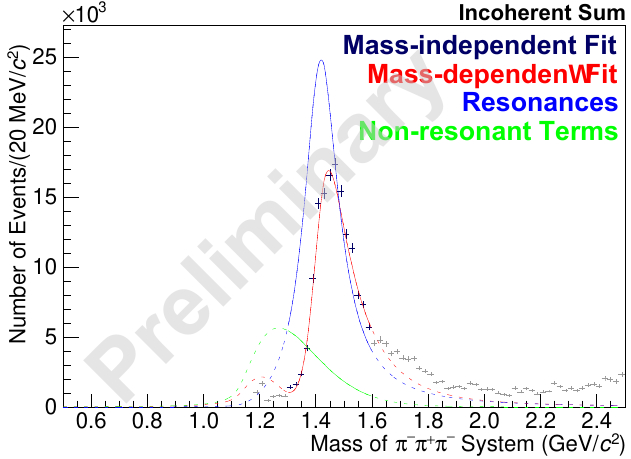}
      \includegraphics[width=0.25\textwidth]{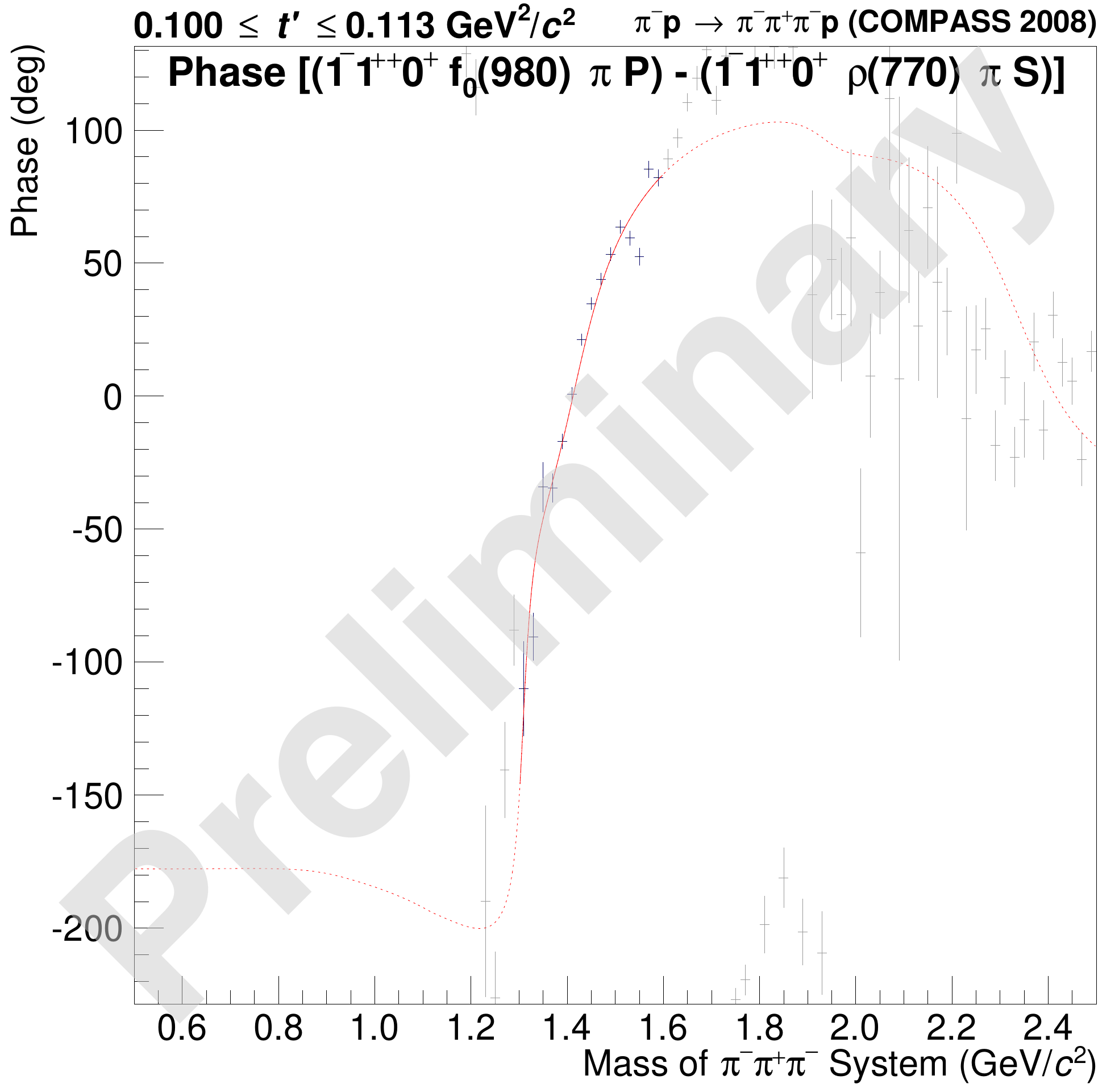}
      \includegraphics[width=0.25\textwidth]{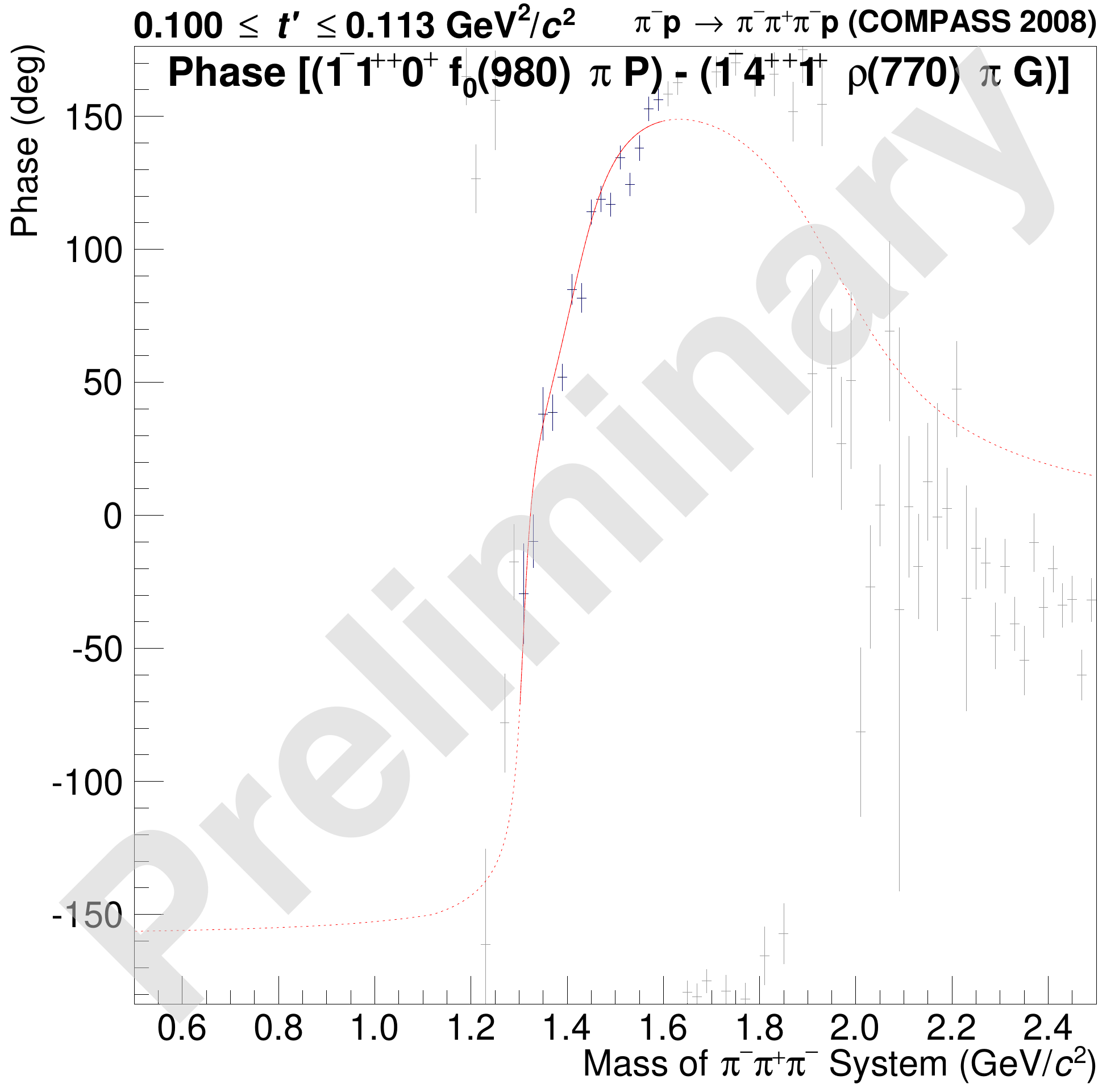}
    \caption{Left: mass spectrum for the $1^{++}0^+ f_0(980)
      \pi P$ wave (Light-grey coloured data points  have not been not used in the mass-dependent fit). Centre: relative phase of the $1^{++}0^+ f_0(980)
      \pi P$ wave with respect to $1^{++}1^+ \rho \pi S$ in the kinematic range $0.100 \leq t'
      \leq 0.113\GEVC$. The model curve is shown in red. Right: Phase relative
      to the $4^{++}1^+ \rho \pi G$ wave.}
    \label{fig:phase_motions_new_a1_low_t}
  \end{center}
  
  \vspace{-0.4cm}
  
\end{figure*}

The main result of  the aforementioned {\it mass-independent fit} is a spin-density matrix of size $80x80$ for each of the 11 bins in $t'$ and 100 bins in mass between 0.5 and 2.5 $\GEV$. In a second step this set of matrices is described by a model which in turn contains assumptions on resonances (typically described by dynamic Breit-Wigner functions) and non-resonant contributions\footnote{non-resonant contributions are described by two functions of the form 
  $\mathcal{F}_{\text{NR}, 1}(m, t') = (m - m_0)^{c_0}\, e^{(c_1 + c_2 t + c_3 t^2)\, q^2}$ for $1^{++}, 2^{++}, 2^{-+}-waves$ and 
  $\mathcal{F}_{\text{NR}, 2}(m) = e^{c_1\, q^2}$ for all other waves.
}. 
We have picked a sub-matrix containing the following waves with $\varepsilon=+1$ and $\textit{M}=0$ and $\textit{M}=1$ for ($a_2$ and $a_4$):
\begin{itemize}
\item $1^{++}0^+ \rho \pi S$ wave: two resonant terms for the $a_1(1260)$ and an $a_1'$
\item $2^{++}1^+ \rho \pi D$ wave: two resonant terms for the $a_2(1320)$ and an $a_2'$
\item $2^{-+}0^+ f_2(1270) \pi S$ wave: two resonant terms for the $\pi_2(1670)$ and a $\pi_2'$
\item $4^{++}1^+ \rho \pi G$ wave: one resonant term for the $a_4(2040)$
\item $1^{++}0^+ f_0(980) \pi P$ wave: one resonant term
\item $0^{-+}0^+ f_0(980) \pi S$ wave: one resonant term for the $\pi(1800)$
\end{itemize}
It has to be noted that the $0^{++}$-isobar is parametrized with the narrow $f_0(980)$ described by a Flatt\'e distribution 
\cite{Flatte:1976}
and a broad structure, 
of which the spectral shape follows 
 a parametrization by \cite{AMP}. 
While production amplitudes and phases of all components can vary with $t'$, resonance parameters do not. Also the shape of the non-resonant contributions has a predetermined $t'$-dependence$^1$. An example for the results of a fit in one bin of $t'$ is shown in fig.~\ref{fig:stamps_mid-t}. 
In detail we show the result for two waves with $J^{PC}=1^{++}$. Fig. \ref{fig:1pp_rho_m0_3t} depicts examples for the $t'$-dependence of the $a_1(1260)$ region. The broad structure is composed of the $a_1(1260)$ resonance interfering constructively (low $t'$) or destructively (high $t'$) with a broad non-resonant component (likely caused by the Deck effect \cite{Deck}). Thus the $t'$-dependence allows for the first time to disentangle two components by their respective and very distinct $t'$-dependence.
The second wave uses the $f_0(980)$ as isobar which is considered as a complex object of likely molecular structure \cite{klempt_08} coupling to both, $\pi\pi$ and $K\overline{K}$. Fig. \ref{fig:phase_motions_new_a1_low_t} depicts the spectral function integrated over \tprim as well as the relative phase motion w.r.t. two different waves. The data exhibit a strong enhancement at a mass around 1420 $\MEV$ connected with phase variation of almost $180^\circ$. No such object has previously been observed.
\section{Analysis of $\pi\pi$~S-wave Structure}
\label{sec-3}
\begin{figure*}[!htb]
  \begin{center}
      \includegraphics[width=0.25\textwidth]{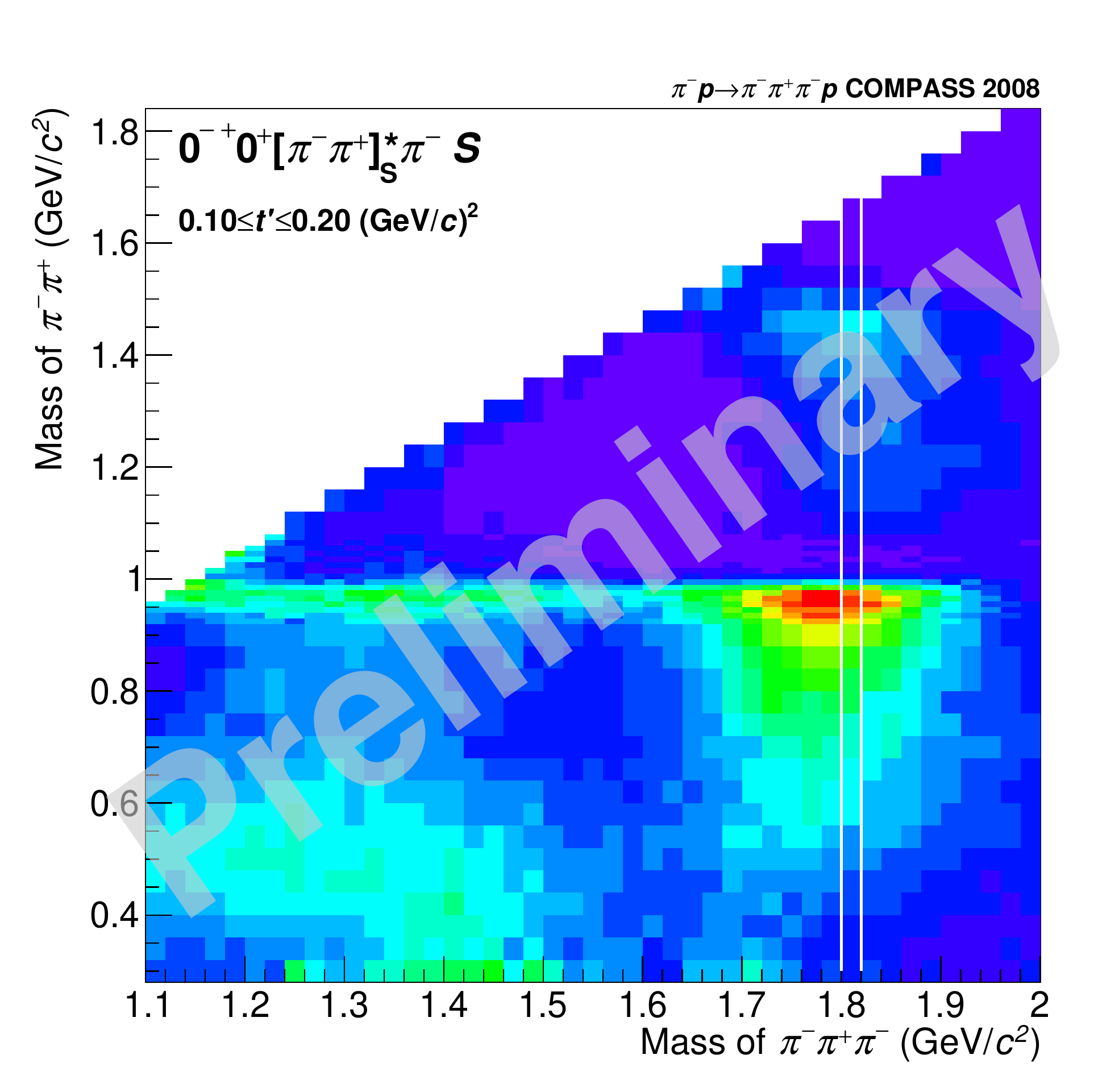}
      \includegraphics[width=0.25\textwidth]{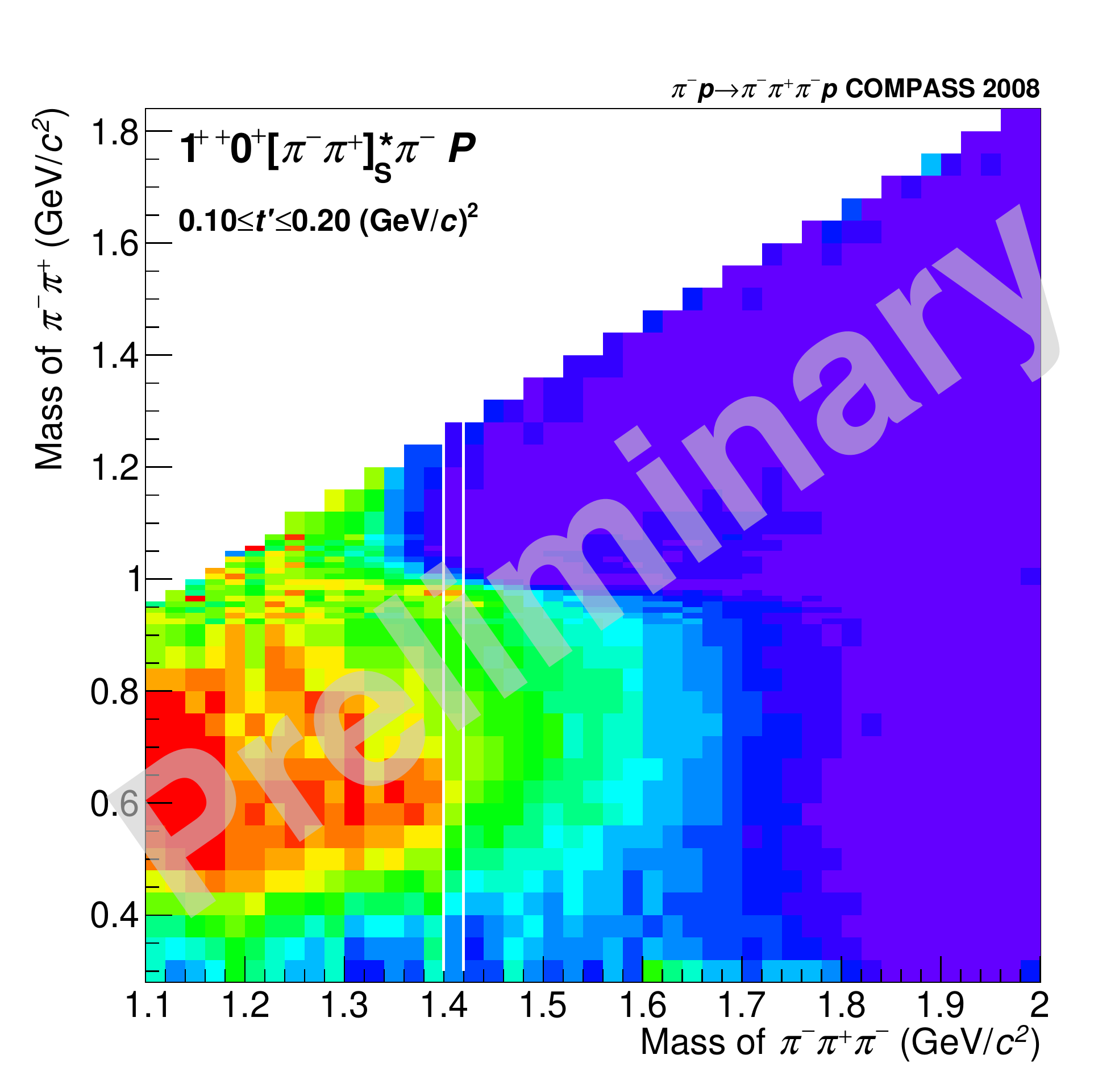}
      \includegraphics[width=0.25\textwidth]{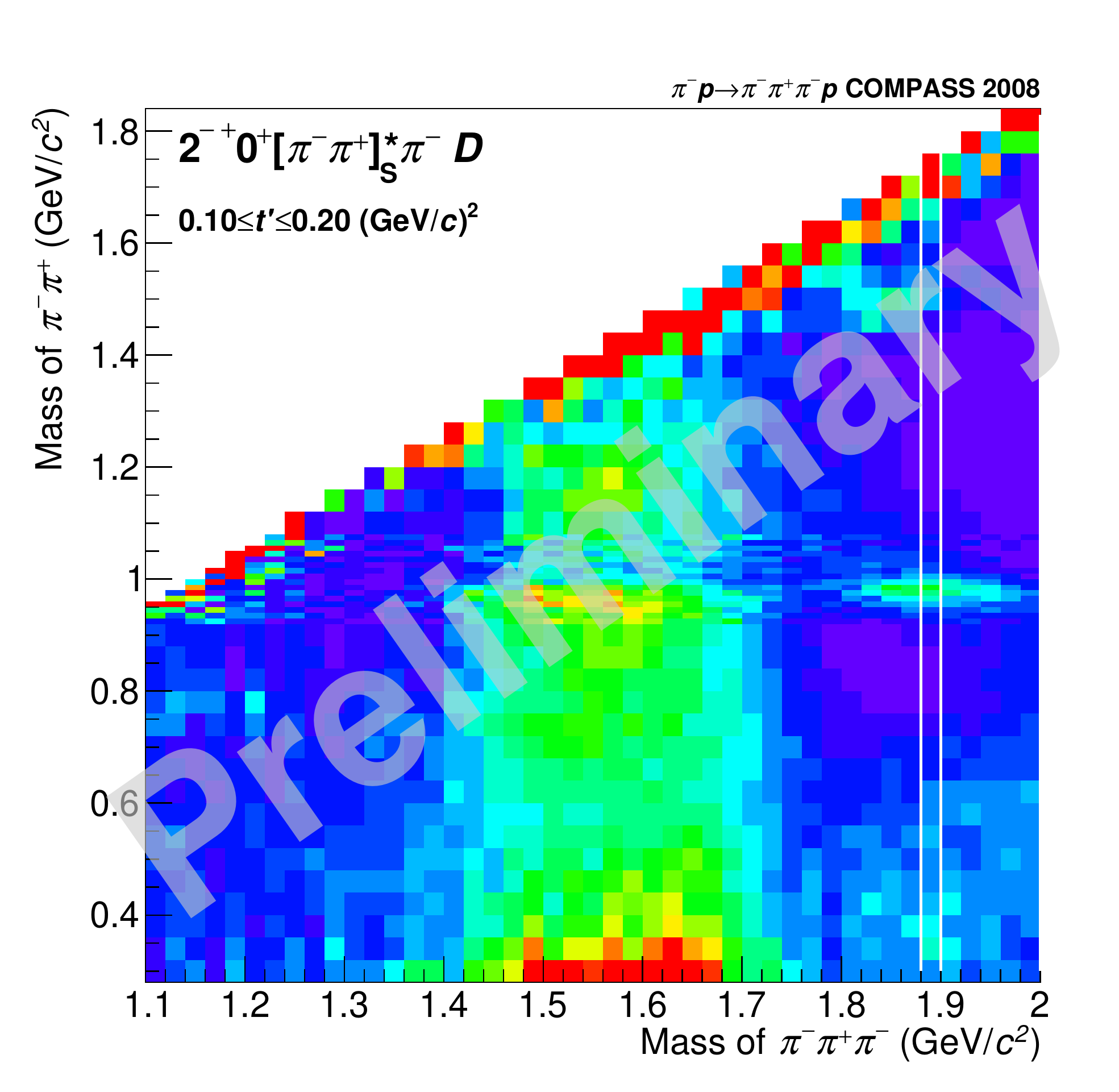}\\
      \includegraphics[width=0.25\textwidth]{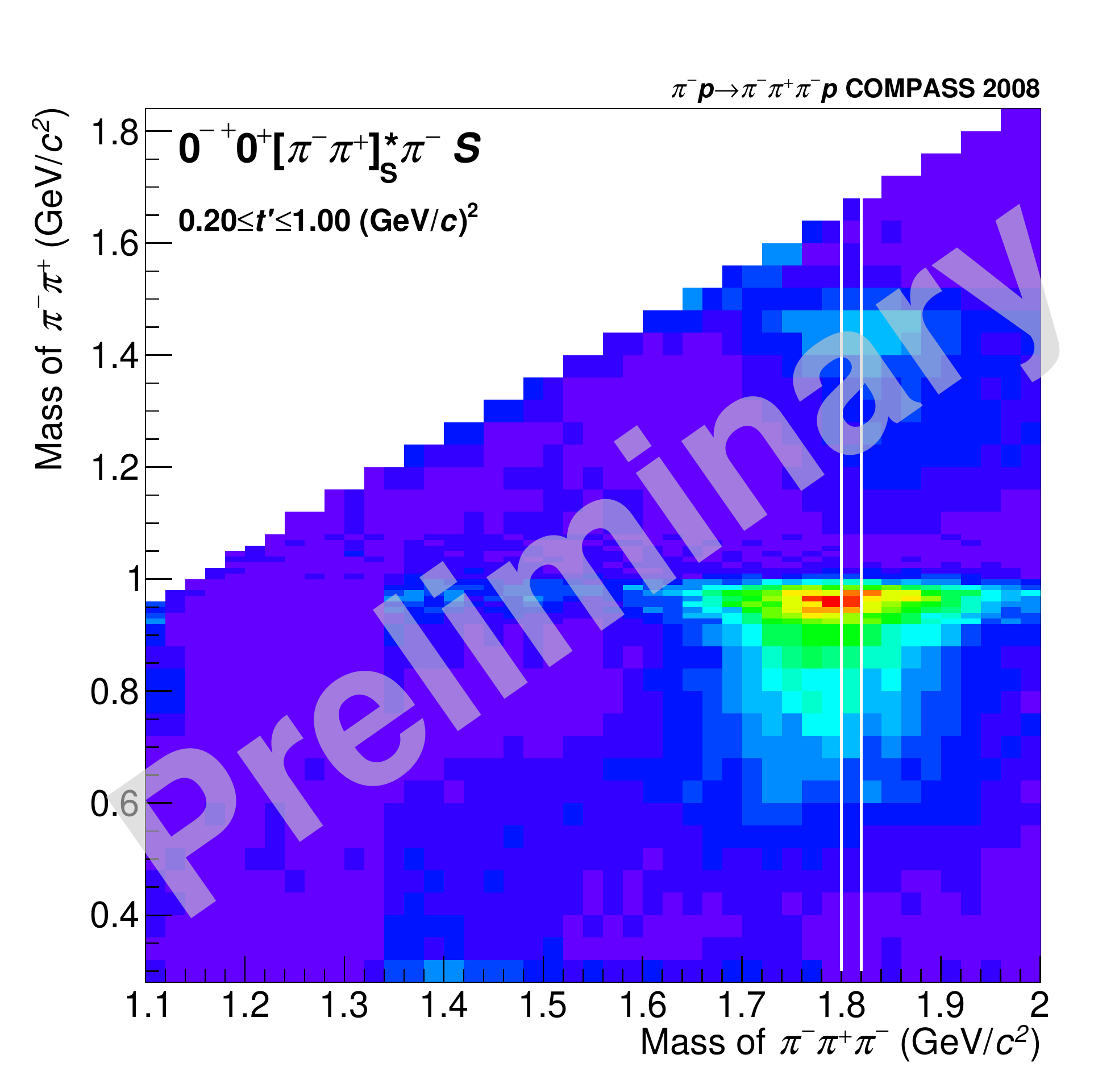}
      \includegraphics[width=0.25\textwidth]{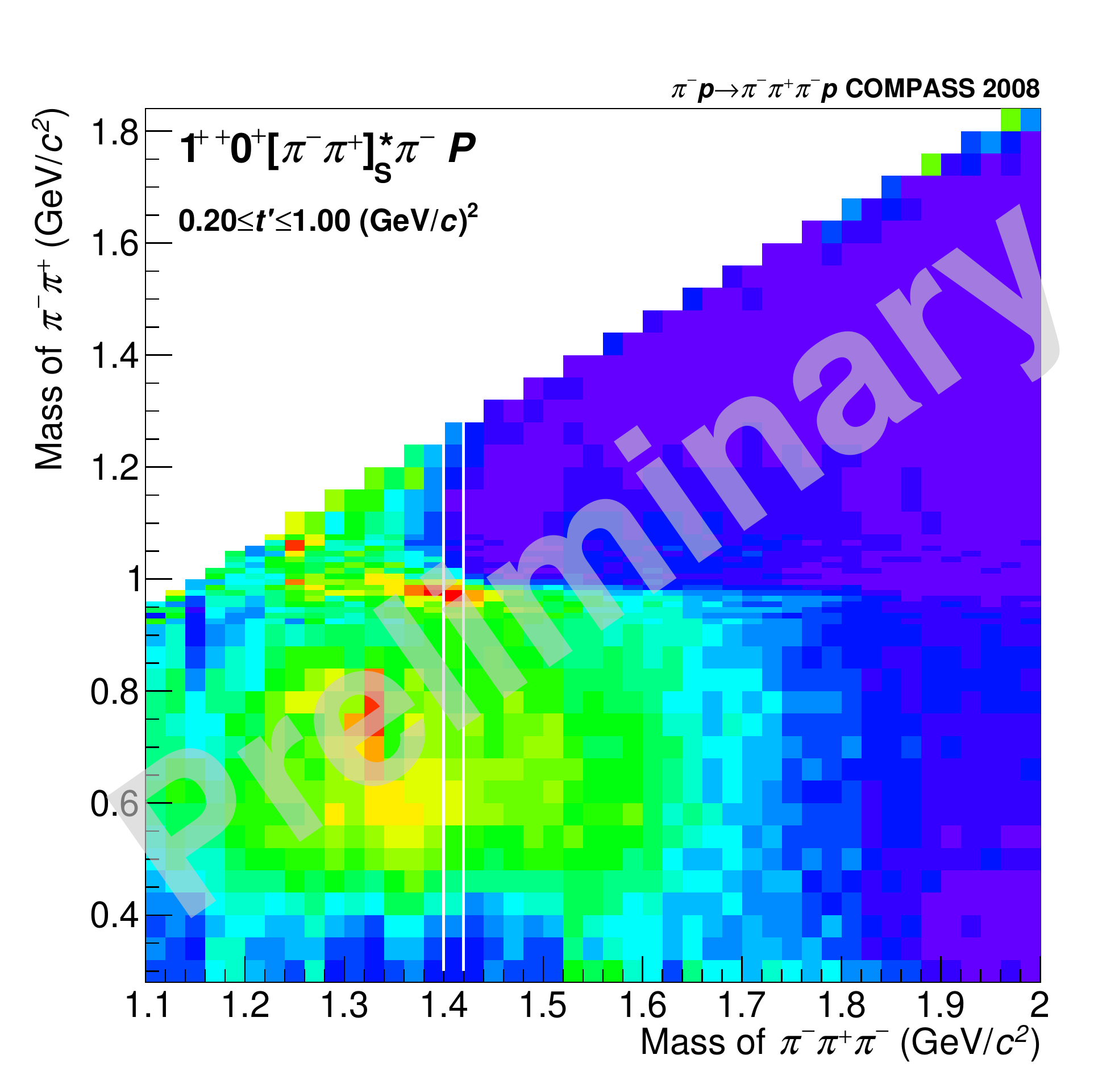}
      \includegraphics[width=0.25\textwidth]{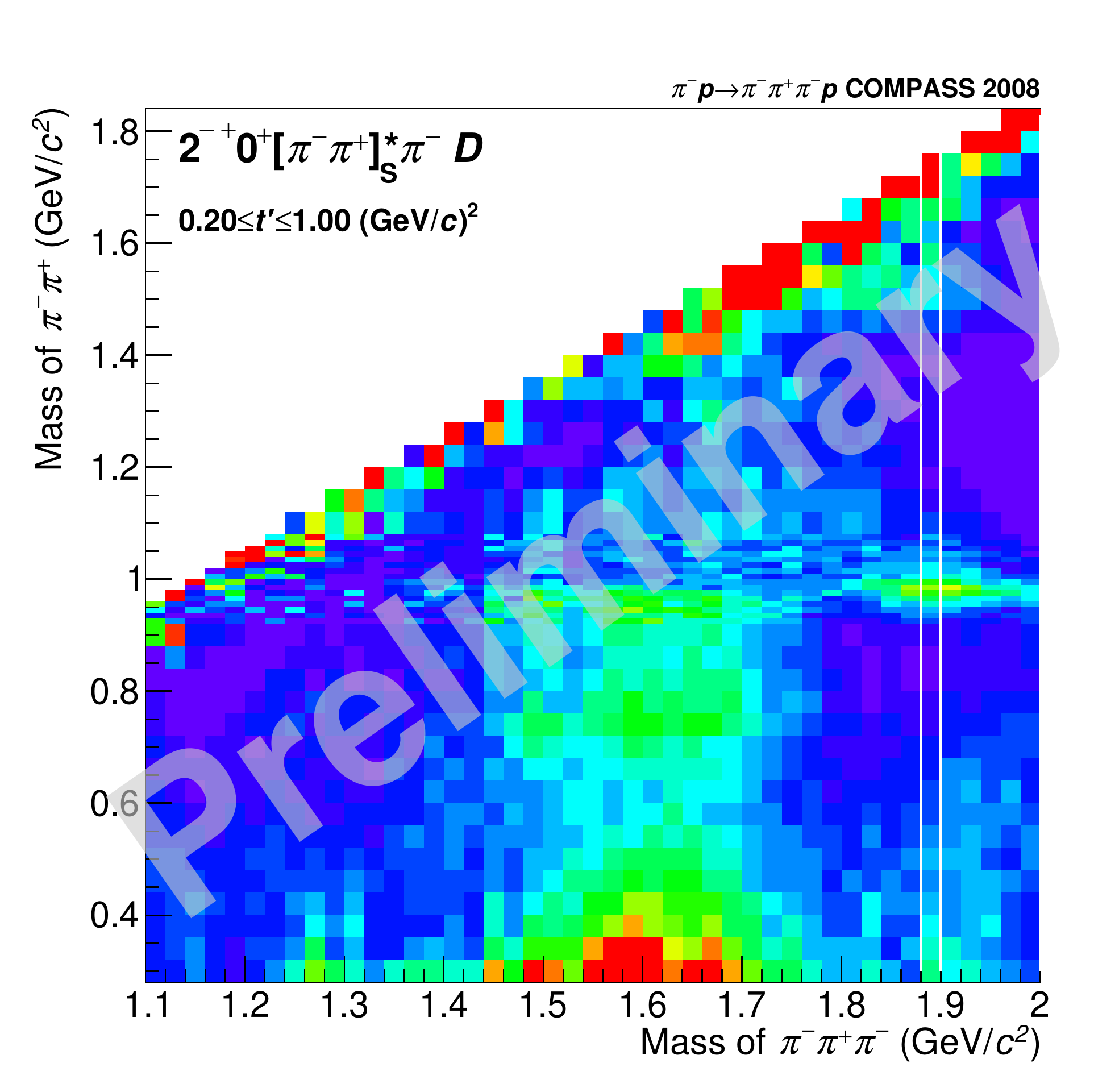}
    \caption{$3\pi$ partial-wave and \PIPIS-isobar correlation for $0^{-+}0^+ \PIPIS \pi S$, $1^{++}0^+ \PIPIS \pi P$ and  $2^{-+}0^+ \PIPIS \pi D$ for low $\tprim\in [0.1,0.2]$ \GEVC and high $\tprim\in [0.2,1.0]$ \GEVC. The central vertical lines indicate the \MPPP mass interval used for the Argand diagrams depicted in fig.~\ref{fig:pipis_argand}.}
    \label{fig:pipis_2D}
  \end{center}
\end{figure*}

\begin{figure*}[!htb]

\vspace{-0.3cm}

  \begin{center}
      \includegraphics[width=0.24\textwidth]{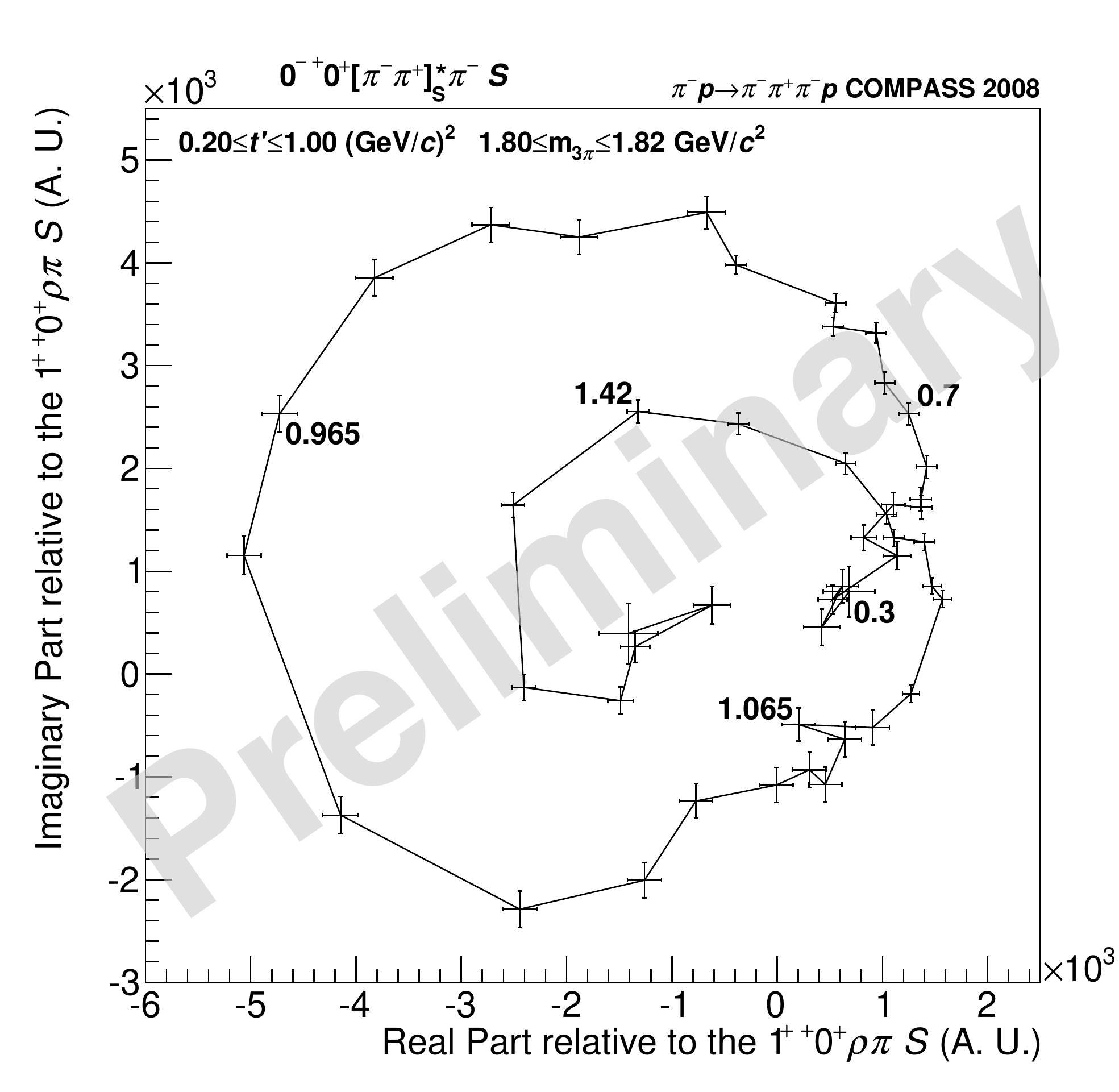}
      \includegraphics[width=0.24\textwidth]{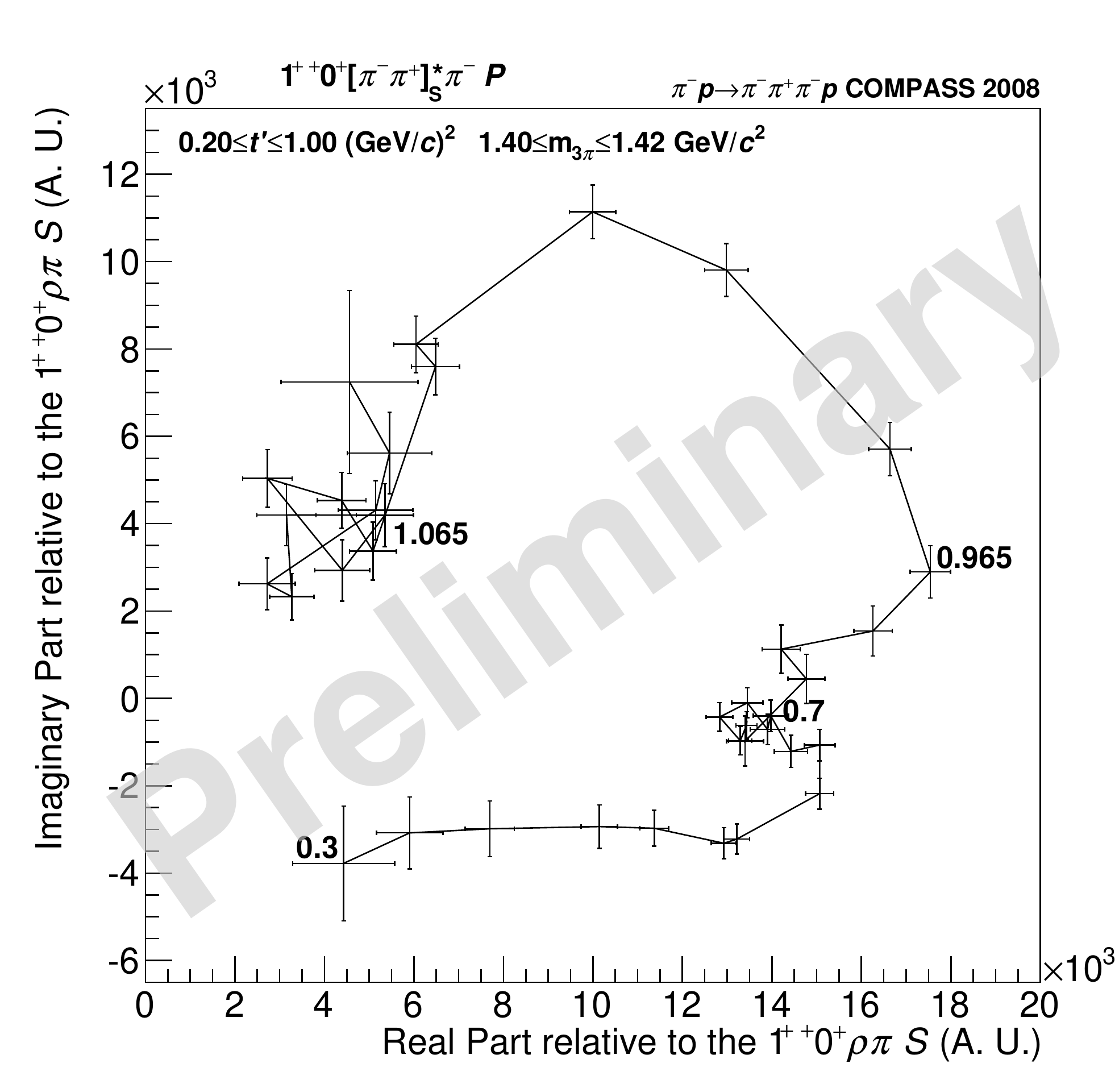}
      \includegraphics[width=0.24\textwidth]{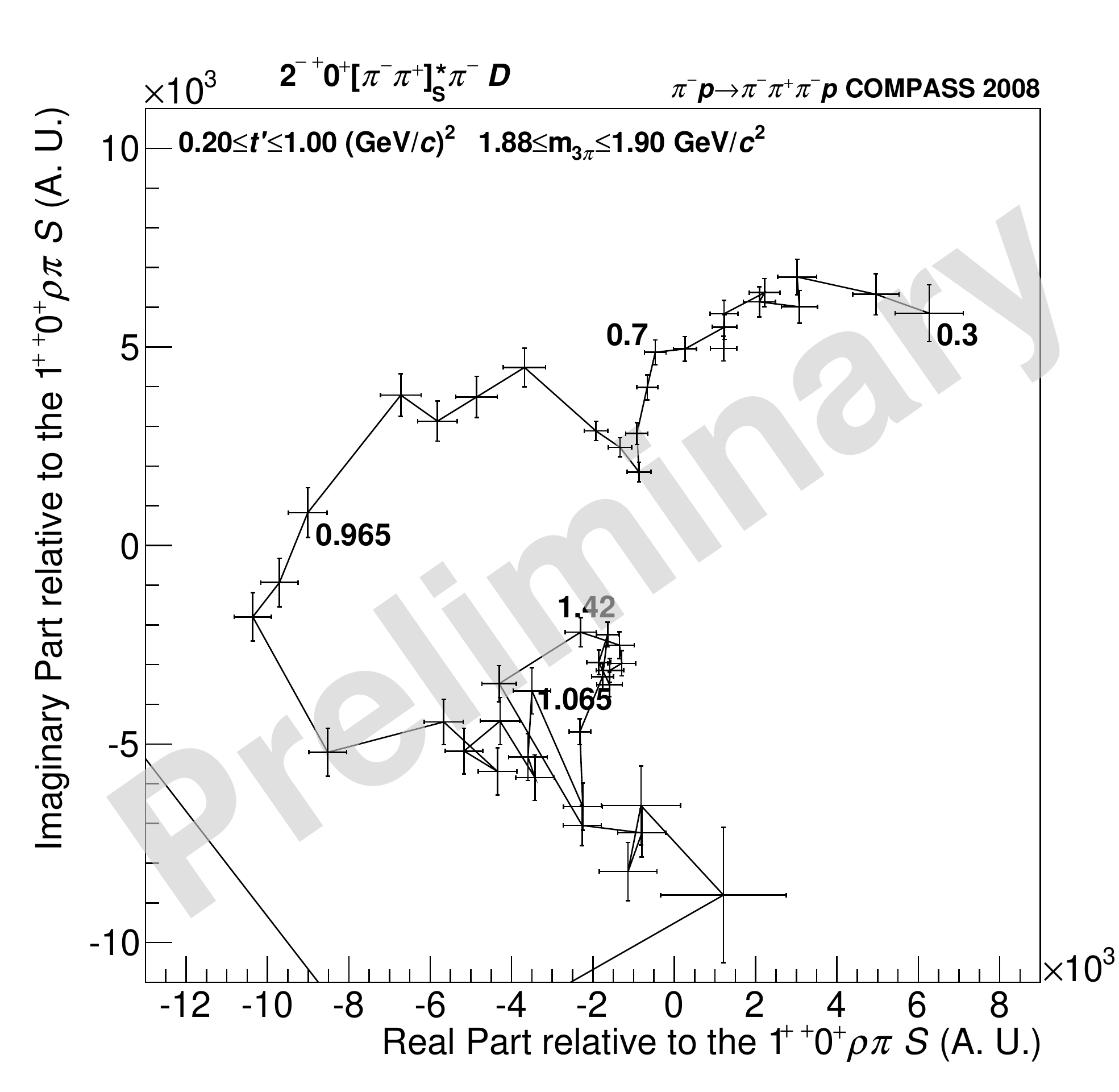}\\
    \caption{Argand diagrams for $\pi^+\pi^-$ $0^{++}$-isobar extracted for $m_{3\pi}$ at the $\pi(1800)$, $a_1(1420)$ and $\pi_2(1880)$ resonances for high $\tprim \in [0.2,1.0]\GEVC$ (see white lines in fig.~\ref{fig:pipis_2D}).}
    \label{fig:pipis_argand}
    
    \vspace{-0.2cm}
    
  \end{center}
\end{figure*}

In order to support the parametrization of the $\pi\pi$~S-wave structure by our model we have developed an alternative PWA-method in which we omit a predetermined parametrization and thus "de-isobar" our $J^{PC}M^{\varepsilon}0^{++}\pi$ wave. This is done for each bin in $m_{3\pi}$ by replacing the functional isobar description discussed above by a series of step-functions, each defined over a small mass bin of 40 \MEV, over a mass range of $[2m_{\pi},m_{3\pi}-m_{\pi}]$. The binning is 10 \MEV around the mass range of the $f_0(980)$. As we steeply increase the effective number of {\it isobars} this fit has been performed in two bins of $t'$ only. The result is depicted in fig. \ref{fig:pipis_2D} for three partial-waves, $0^{-+}, 1^{++}$ and $2^{-+}$ for the $3\pi$-system.
In addition to the mass spectra we also extract the Argand diagrams which are depicted in fig.~\ref{fig:pipis_argand}  for $\tprim \in [0.2,1.0]\GEVC$, in the mass region of the resonances observed, namely $\pi(1800)$, $a_1(1420)$ and $\pi_2(1880)$. It should be noted that the phases are measured w.r.t. the $1^{++}0^+\rho\pi S$ wave and contain the sum of both, production and decay phases. The strong dependence of the distributions in fig.~\ref{fig:pipis_2D} on $t'$ is striking and reveals that much of the structure observed in the unresolved distributions is due to non-resonant processes with steeply falling $t'$ dependence. Thus we clearly disentangle resonant and non-resonant components and identify the dominant role of the iso-scalar resonances $f_0(980)$ and $f_0(1500)$ in the decay process of the three $3\pi$-states (pair of white lines in fig.~\ref{fig:pipis_argand}). The resonance structure is also reflected in the circles observed in the Argand diagrams. Here, the absence of phase motion in the $\pi\pi$-System related to $a_1(1420)$ is very distinct (no circle in the Argand diagram visible) and hints to production and decay phase having opposite sign and similar magnitude. 
We thus conclude that the $a_1(1420)$ coupling to $f_0(980)$ is genuine and not an artifact of the isobar parametrization and no strong coupling to the broad component in $\pi\pi$~S-wave channel can be observed. More details, however, require a mass-dependent fit relating the $\pi^+\pi^-$ $0^{++}$ mass and phase spectra to the $3\pi$ systems.
\section{Conclusions}
\label{sec-4}

\vspace{-2mm}

Using the worlds largest data set on diffractive pion dissociation we have developed new analysis tools. This allows to disentangle resonant and non-resonant processes contributing to multi-particle production using the reaction \Reaction. We have extracted the $a_1(1260)$ resonance and observed hints for excited states of the $a_1$ and $a_2$ around $1800-1900 \MEV$. In particular we have observed a new iso-vector meson $a_1(1420)$ decaying into $f_0(980)\pi$ with no clear signs of concurring decay modes. A full phase motion of $180^\circ$ characteristic to resonance production has been observed. The origin of this object with a width of about $140 \MEV$ is yet unclear. The similarity in mass and width to the $f_1(1420)$ strongly coupling to $KK^*$ is striking. At present we cannot exclude re-scattering effects involving a coupling of $KK^*$ and $f_0(980)\pi$ as outlined in \cite{Berger}. A detailed coupled channel analysis including the study of the $K\overline{K}$ final state would be necessary. In addition we have studied the iso-scalar structure of the $\pi\pi$ isobar in the decay of $J^{PC}=0^{-+}, 1^{++}, 2^{-+}$. A clear coupling of $a_1(1420)$ to $f_0(980)$ and of $\pi(1800)$ and $\pi_2(1880)$ to both, $f_0(980)$ and $f_0(1500)$, has been observed in addition to a broad $\pi\pi$~S-wave component, exhibiting a strong $t'$ dependence .

%
%
%

\end{document}